\documentclass[amsmath,amssymb,superscriptaddress,prl,twocolumn,showkeys]{revtex4-2}

\usepackage{graphicx}
\usepackage{siunitx}
\usepackage{pdfpages}
\usepackage{hyperref}

\hypersetup{colorlinks=true,citecolor=blue,linkcolor=blue,urlcolor=blue}

\newcommand{\uJcm}{\si{\micro\joule\per\centi\metre\squared}}
\newcommand{\cm}{\si{\per\centi\metre}}
\newcommand{\Ag}{$A_{\mathrm{g}}$}
\newcommand{\VSe}{1\textit{T}-VSe$_{2}$}
\newcommand{\Fc}{$F_c$}

\newcommand{\Tel}{$T_\mathrm{el}$}
\newcommand{\Tph}{$T_\mathrm{ph}$}
\newcommand{\delR}{$\Delta R/R$}
\newcommand{\qCDW}{$q_{\mathrm{CDW}}$}
\newcommand{\TCDW}{$T_{\mathrm{CDW}}$}
\newcommand{\GammaBar}{$\bar{\Gamma}$}
\newcommand{\MBar}{$\bar{\mathrm{M}}$}
\newcommand{\KBar}{$\bar{\mathrm{K}}$}

\newcommand{\ABar}{$\bar{\mathrm{A}}$}
\newcommand{\LBar}{$\bar{\mathrm{L}}$}

\makeatletter
\AtBeginDocument{\let\LS@rot\@undefined}
\makeatother

\begin{document}
	

\title{Anomalous amplitude mode dynamics below the expected charge-density-wave transition in 1\textit{T}-VSe$_{2}$}

	
\author{Charles J. Sayers}
\email[Correspondence email address: ]{charles.sayers@polimi.it}
\affiliation{Dipartimento di Fisica, Politecnico di Milano, 20133 Milano, Italy} 
	
\author{Giovanni Marini}
\affiliation{Department of Physics, University of Trento, Via Sommarive 14, 38123 Povo, Italy}
\affiliation{Graphene Labs, Fondazione Istituto Italiano di Tecnologia, Via Morego, I-16163 Genova, Italy}
	
\author{Matteo Calandra}
\affiliation{Department of Physics, University of Trento, Via Sommarive 14, 38123 Povo, Italy}
\affiliation{Graphene Labs, Fondazione Istituto Italiano di Tecnologia, Via Morego, I-16163 Genova, Italy}
	
\author{Hamoon Hedayat}
\affiliation{University of Cologne, Institute of Physics II, 50937 Cologne, Germany}
	
\author{Xuanbo Feng}
\affiliation{Institute of Physics, University of Amsterdam, Science park 904, 1098 XH Amsterdam, The Netherlands}
	
\author{Erik van Heumen}
\affiliation{Institute of Physics, University of Amsterdam, Science park 904, 1098 XH Amsterdam, The Netherlands}
	
\author{Christoph Gadermaier}
\affiliation{Dipartimento di Fisica, Politecnico di Milano, 20133 Milano, Italy}
	
\author{Stefano Dal Conte}
\affiliation{Dipartimento di Fisica, Politecnico di Milano, 20133 Milano, Italy}
	
\author{Giulio Cerullo}
\email[Correspondence email address: ]{giulio.cerullo@polimi.it}
\affiliation{Dipartimento di Fisica, Politecnico di Milano, 20133 Milano, Italy}


\begin{abstract}

A charge-density-wave (CDW) is characterized by a dynamical order parameter consisting of a time-dependent amplitude and phase, which manifest as optically-active collective modes of the CDW phase. Studying the behaviour of such collective modes in the time-domain, and their coupling with electronic and lattice order, provides important insight into the underlying mechanisms behind CDW formation. In this work, we report on femtosecond broadband transient reflectivity experiments on bulk \VSe~using near-infrared excitation. At low temperature, we observe coherent oscillations associated with the CDW amplitude mode and phonons of the distorted lattice. Across the expected transition temperature at 110 K, we confirm signatures of a rearrangement of the electronic structure evident in the quasiparticle dynamics. However, we find that the amplitude mode instead softens to zero frequency at 80 K, possibly indicating an additional phase transition at this temperature. In addition, we demonstrate photoinduced CDW melting, associated with a collapse of the electronic and lattice order, which occurs at moderate excitation densities, consistent with a dominant electron-phonon CDW mechanism.
   
\end{abstract}


	
	
\maketitle

	
\section*{Introduction}
	
The metallic transition metal dichalcogenides (TMDs) are layered materials which host various strongly-correlated phases of matter, but are arguably best known for exhibiting charge-density-waves (CDWs) \cite{Wilson1975}. Despite decades of research, the search for a suitable microscopic description to explain CDWs in real systems is an ongoing goal of condensed matter physics \cite{Zhu2015}. The group-5 TMDs, i.e. with transition metal V, Nb or Ta, are prototypical CDW systems, which as a result of their quasi-2D Fermi surfaces, have been described by Fermi surface nesting (FSN) \cite{Wilson1975,Aebi2001}, a higher-dimensional manifestation of the original theory by Peierls in the 1950s for a 1D system \cite{Peierls1955}. However, Johannes and Mazin cast doubt on the role of FSN in any real system with dimensionality greater than 1D, or in the presence of small perturbations \cite{JohannesMazin2008}. Rather, a momentum-dependent electron-phonon coupling (EPC) is more likely to be significant in the majority of TMDs \cite{Rossnagel2011,Henke2020}.
	 
\VSe~is a group-5 TMD that has gathered significant attention because of the dramatic change in its electronic properties and emergent phenomena in the monolayer limit, including; strongly enhanced CDW order \cite{Chen2018,Feng2018}, a metal-to-insulator transition at high temperature \cite{Duvjir2018}, and possible room-temperature ferromagnetism \cite{Ma2012,Bonilla2018} which remains controversial \cite{Feng2018,Fumega2019}. A zoo of charge-ordered phases with various symmetries have been found \cite{Chen2018,Chen2022,Chua2022}, and linked to substrate interactions, opening the possibility of strain engineering \cite{Zhang2017}. In van der Waals heterostructures containing \VSe, signatures of band hybridization and strong correlations have been observed in the form of flat bands \cite{Yilmaz2021} and spectral kinks \cite{Yilmaz2022}.
	
Intense ultrashort laser pulses have been used to study the dynamics of electronic and lattice order in the time-domain for a range of strongly correlated materials, including CDW systems \cite{Demsar1999,Eichberger2010,Peterson2011,Porer2014,Hedayat2021}, Mott insulators \cite{Perfetti2006,Sohrt2014,Ligges2018} and superconductors \cite{Kabanov1999,Gadermaier2010,DalConte2015,Boschini2018}. Aside from clocking the timescale of relevant quasiparticle dynamics to learn about the underlying physical mechanisms in such systems \cite{Hellmann2012}, time-domain experiments benefit from direct access to collective modes of the ordered state. The order parameter of the CDW phase is a complex quantity $\Psi = |\Delta| e^{i\varphi}$ consisting of a time-dependent amplitude, $|\Delta(t)|$ and phase, $\varphi(t)$ which manifest as dynamical collective excitations of the CDW condensate known as the amplitude (AM) and phase (PM) modes \cite{Gruner1988}, respectively. These modes are optically-active and can be accessed by either; driving a coherent oscillation of the charge density (AM), or inducing the collective sliding of the charge modulation (PM). The AM has been observed in numerous CDW systems, appearing as a temporal modulation of the optical properties following excitation \cite{Demsar1999,Mohr-Vorobeva2011,Porer2014,Werdehausen2018,Sayers2022} or as bandgap oscillations in time-resolved photoemission (tr-ARPES) experiments \cite{Hedayat2019,Sayers2020b}. Evidence for the PM has instead been reported in cuprates using transient grating spectroscopy \cite{Torchinsky2013} or more recently in 1D systems by detection of THz emission \cite{Kim2023}. 

In \VSe, indirect evidence for the PM has been provided by infrared optical spectroscopy \cite{Feng2021}, while limited evidence for the AM comes from Raman experiments \cite{Sugai1985}. Recent time-domain experiments using tr-ARPES \cite{Biswas2021,Majchrzak2021,Bronsch2024} investigated the electron dynamics of bulk \VSe~but only at relatively high laser fluence of several mJ cm$^{-2}$, where they found no evidence for coherent phonons or the AM of the CDW phase.
	
Here, we report on femtosecond broadband transient reflectivity experiments on bulk \VSe~using near-infrared (1.55 eV) excitation to investigate the collective mode dynamics of the CDW phase. In the non-perturbative (low fluence) regime, and at low temperatures, we directly observe coherent oscillations in the time-domain which we assign to the AM. A detailed temperature dependence allows us to identify signatures of an electronic structure rearrangement at the expected normal-to-CDW transition temperature \TCDW~= 110 K. By contrast, we find that the AM softens towards lower temperatures at $T_{\mathrm{AM}}$ = 80 K. This is further supported by fluence-dependent measurements which demonstrate a photoinduced transition consisting of AM suppression at a lower fluence threshold, followed by electronic and lattice reconstruction at a higher threshold. We find a relatively high threshold fluence of \Fc~= 360 \uJcm~for the photoinduced CDW-to-normal phase transition, inconsistent with a purely FSN scenario in \VSe. We discuss our results of the anomalous AM behaviour in the context of charge-lattice commensurability (incommensurate-to-commensurate CDW) or domain related phenomena (3$q$-to-2$q$ transition) occurring at 80 K, and emphasize the need for further experimental investigation of this topic.

	
\section*{Results \& Discussion}

\subsection{CDW characteristics in equilibrium}
	
\begin{figure}
    \includegraphics[width=1.0\linewidth,clip]{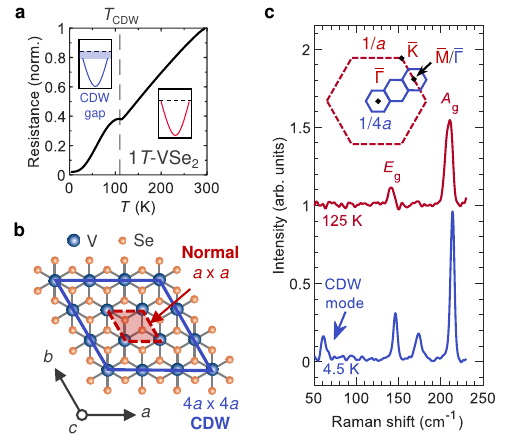}
    \caption{\label{fig:intro} \textbf{Signatures of the equilibrium CDW transition in \VSe.} (\textbf{a}) Electrical resistance as a function of temperature showing the normal-to-CDW phase transition at \TCDW~$\approx$ 110 K. The transition results in the opening of a bandgap along the \KBar-\MBar-\KBar~direction, as illustrated by the inset panels. (\textbf{b}) In-plane view of the crystal structure. The primitive unit cells of the normal phase and reconstructed ($4a \times 4a$)~CDW phase are represented by the red and blue colours, respectively. (\textbf{c}) Raman spectra in the normal (125 K) and CDW phase (4.5 K). The mode symmetries of the normal phase structure are labelled. The inset shows a sketch of the 1/$a$ and 1/4$a$ BZ in each phase, highlighting the equivalent \MBar- and \GammaBar-points in the CDW phase due to the reconstruction.}
\end{figure}

Bulk \VSe~undergoes a CDW transition at \TCDW~$\approx$ 110 K, resulting in a rearrangement of the electronic structure and a small bandgap of $\sim$ 10 meV \cite{Feng2021} opening on portions of the V-3$d$ electron pockets centered around the \MBar~points of the two-dimensional Brillouin zone (BZ) \cite{Terashima2003,Chen2018}, although significant $k_z$ dispersion should also be considered \cite{Sato2004,Strocov2012}. As shown in Fig.~\ref{fig:intro}\textbf{a}, this manifests as a slight increase in the electrical resistance near \TCDW~due to the partially gapped Fermi surface, before exhibiting overall metallic behaviour as the temperature is reduced further \cite{Sayers2020a}. The transition is accompanied by a lattice reconstruction which results in a new in-plane unit cell of $4a \times 4a$~\cite{Tsutsumi1982} as illustrated in Fig.~\ref{fig:intro}\textbf{b}. The corresponding BZ is hence reduced to 1/4$a$ in reciprocal space (see inset Fig.~\ref{fig:intro}\textbf{c}) meaning that the \GammaBar- and \MBar-point become equivalent, resulting in folding of both the electronic band structure and phonon dispersion from \MBar~to~\GammaBar. One important consequence is the appearance of new Raman-active (\GammaBar-point) phonon modes in the reconstructed phase below \TCDW~\cite{Sugai1985}. In addition, there is the onset of the AM and PM which are Raman- and infrared-active, respectively \cite{Torchinsky2013}. In Fig.~\ref{fig:intro}\textbf{c}, we compare Raman spectra of bulk \VSe~measured above and below \TCDW, demonstrating the appearance of new CDW modes.
 
\subsection{Transient response in the CDW phase}
	
\begin{figure}
	\includegraphics[width=1.0\linewidth,clip]{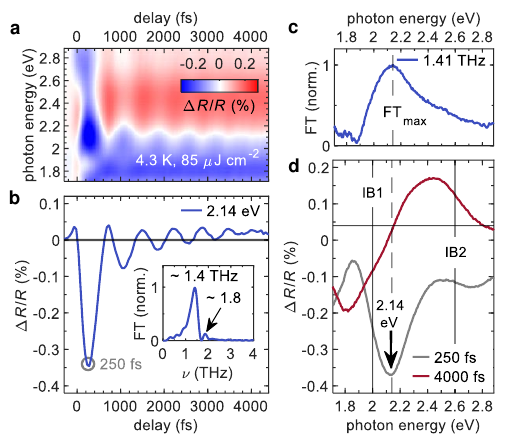}
	\caption{\label{fig:response} \textbf{Transient response for low pump fluence deep in the CDW phase}. (\textbf{a}) Temporal evolution of the differential reflectivity (\delR) for low fluence (85 \uJcm) at $T$ = 4.3 K. (\textbf{b}) Dynamics extracted at 2.14 eV probe photon energy. The inset shows the Fourier transform (FT) of the isolated oscillatory component with two resolvable frequencies labelled. (\textbf{c}) Probe photon energy dependence of the FT magnitude extracted at 1.41 THz. The dashed line indicates the maximum at FT$_{\mathrm{max}}$ $\approx$ 2.14 eV. (\textbf{d}) Differential spectra (\delR) at the maximum transient signal (250 fs) and 4000 fs delay after photoexcitation. The vertical arrow highlights the negative feature in \delR~at 2.14 eV, which also coincides with FT$_{\mathrm{max}}$. The solid vertical lines correspond to interband transitions IB1 (2.0 eV) and IB2 (2.6 eV) reported in Ref.~\cite{Feng2021}.}
\end{figure}

We start by investigating the transient response of \VSe~using 1.55 eV excitation for low pump fluence (85 \uJcm), deep in the CDW phase at $T$ = 4.3 K, as reported in Fig.~\ref{fig:response}\textbf{a}. Notably, the response is dominated by coherent oscillations which are visible across the majority of the measured spectral range $\sim$ 1.7 - 2.9 eV. Dynamics extracted at 2.14 eV, as shown in Fig.~\ref{fig:response}\textbf{b}, reveal oscillations that persist above the noise for at least 5 periods and a total of $\sim$ 4000 fs. Fourier transform (FT) analysis of the isolated oscillatory component in the inset of Fig.~\ref{fig:response}\textbf{b}, shows a broad peak centered at $\sim$ 1.4 THz (47 \cm) and a weaker component on the high frequency side at $\sim$ 1.8 THz (60 \cm).

\begin{figure}
    \includegraphics[width=1.0\linewidth,clip]{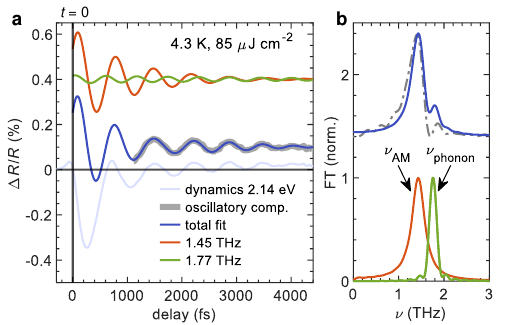}
    \caption{\label{fig:frequency} \textbf{Frequency analysis of the CDW oscillations}. (\textbf{a}) Isolated oscillatory component (grey) extracted from the measured dynamics at 2.14 eV (light blue) for low fluence (85 \uJcm) at $T$ = 4.3 K. The oscillatory component is fitted with the sum of two damped sinusoidal functions (dark blue) with frequencies $\sim$ 1.45 THz (orange) and 1.77 THz (green). Curves are offset vertically for clarity. At $t$ = 0 (vertical line), both modes exhibit an oscillatory maximum, confirming their cosine nature. (\textbf{b}) Fourier transform (FT) of the individual frequency components of the fit  (green and orange) and the total (blue) in panel \textbf{a}. FT of the measured data is shown for comparison (grey dashed line). The low- and high-frequency components are identified as the AM, and a phonon of the CDW lattice, respectively.}
\end{figure}

Our broadband probe allows us to spectrally resolve both the oscillation amplitude and phase, which is important in understanding the link between the coherent response and excitation of interband transitions in CDW materials \cite{Sayers2022}. Fig.~\ref{fig:response}\textbf{c} shows the probe photon energy dependence of the dominant mode at $\nu \approx$ 1.4 THz from FT analysis, which displays a clear maximum near 2.14 eV. Interestingly, this energy corresponds precisely to the short-lived negative feature visible in the \delR~map in Fig.~\ref{fig:response}\textbf{a} within the first few hundreds of femtoseconds and highlighted in the transient spectrum (250 fs) in Fig.~\ref{fig:response}\textbf{d}. A detailed temperature dependence of the transient spectra (see Supporting Information A) shows that the negative peak, initially at 2.14 eV, weakens and redshifts with increasing temperature before disappearing for $T >$ \TCDW, suggesting that this feature is closely related to the CDW phase. A further inspection of the transient spectra in Fig.~\ref{fig:response}\textbf{d} shows that the 2.14 eV feature is actually the low energy minimum of a derivative shape with its centre at $\sim$ 2.0 eV, while another can be seen centred around $\sim$ 2.6 eV. These energies are in excellent agreement with interband (IB) transitions found in this spectral region by steady-state optical spectroscopy \cite{Feng2021}, and labelled IB1 and IB2 respectively in Fig.~\ref{fig:response}\textbf{d}. We also observe a $\pi$ phase shift of the oscillations (directly visible in Fig.~\ref{fig:response}\textbf{a}) across $\sim$ 1.9 eV, corresponding reasonably well to IB1, which also explains the sharp drop towards zero oscillatory magnitude at this energy in Fig.~\ref{fig:response}\textbf{c}, as observed in other systems across known optical transitions \cite{Sayers2022,Sayers2023b,Mor2021}.

Analysis of the isolated oscillatory component in Fig.~\ref{fig:frequency}\textbf{a} shows that the total coherent response is well described by the sum of two damped oscillators according to:

\begin{equation}\label{eq:cosinedamp}
\begin{split}
        \Delta R/R (t) = A_{1}e^{-t/\tau_{\mathrm{d1}}}~\mathrm{cos}(2\pi\nu_{1} t + \phi_{1}) \\ + A_{2}e^{-t/\tau_{\mathrm{d2}}}~\mathrm{cos}(2\pi\nu_{2} t + \phi_{2})
\end{split}
\end{equation}

\noindent
By fitting the data with Equation~\ref{eq:cosinedamp}, we find an initially intense mode with frequency $\nu_{1}$ = 1.45 THz (48 \cm) which is strongly damped ($\tau_{\mathrm{d1}}$ = 940 fs), and a less intense component at $\nu_{2}$ = 1.77 THz (59 \cm) which is more weakly damped ($\tau_{\mathrm{d2}}$ = 3990 fs). These frequencies are in good agreement with the Raman-active \Ag-symmetry modes which appear in the CDW phase as a pair of modes around 50 and 62 \cm~as observed in previous Raman studies~\cite{Sugai1985}, although here they are slightly red-shifted, presumably due to the effect of weak photoexcitation. In those works, the lowest frequency mode was identified as the AM based on its strong temperature-dependence, while the higher frequency mode is likely a phonon of the distorted lattice \cite{Sugai1985}. This assignment is consistent with the FT analysis in Fig.~\ref{fig:response}\textbf{c} as we find that the $\sim$ 1.4 THz mode strongly modulates the optical response in the vicinity of interband transitions (IB1) associated with the reconstructed phase \cite{Sayers2022}, indicative of the strong coupling with the electronic structure expected for the AM \cite{Sayers2020b,Sayers2022APS,Ruggeri2024}.

The individual frequency components of the fit are shown in Fig.~\ref{fig:frequency}\textbf{a}, and reveal oscillatory maxima, within the temporal resolution of the experiment, at zero delay ($t$ = 0) for both modes, i.e. a cosine phase $\phi_{1} = \phi_{2}$ = 0, typically associated with the displacive excitation of coherent phonons (DECP) \cite{Zeiger1992}, and in good agreement with their expected \Ag~symmetry. As mentioned previously, we observe a phase flip associated with IB1, such that for all measured photon energies above or below $\sim$1.90 eV we find a cosine phase $\phi_{1} = \phi_{2}$ = 0 or $\pi$, respectively. To visualize the result in the frequency domain, we also performed FT analysis of all components of the fit, as shown in Fig.~\ref{fig:frequency}\textbf{b}, which shows a good match between the total fit (blue) and raw data (grey), validating our approach.

Finally, we note that the relatively short lifetime of the coherent response is similar to other group-5 TMDs with weak CDW order such as 2$H$-NbSe$_2$ (\TCDW = 33 K) \cite{Xi2015,Anikin2020}, or 2$H$-TaSe$_2$ (\TCDW = 122 K) in the incommensurate phase \cite{Demsar2002}. Whereas, in other systems with strong commensurate CDW order, such as 1$T$-TaSe$_2$ (\TCDW = 473 K), the lifetime is significantly longer $\tau_{\mathrm{d}} \approx$ 6 ps (at 80 K) and the oscillations persist for tens of picoseconds \cite{Sayers2022}.

\subsection{Dynamics across the thermal CDW transition}

\begin{figure}
    \includegraphics[width=1.0\linewidth,clip]{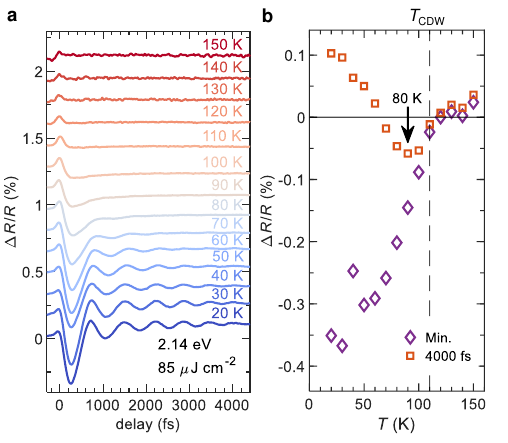}
    \caption{\label{fig:dynamics} \textbf{Quasiparticle dynamics across the thermal CDW transition}. (\textbf{a}) Temperature dependence ($T$ = 20 - 150 K) of the dynamics extracted at 2.14 eV for low fluence (85 \uJcm). Data are offset for clarity. (\textbf{b}) Magnitude of the transient signal, \delR~at its local minimum (min.) and 4000 fs as a function of temperature showing a significant rearrangement at \TCDW~= 110 K (vertical dashed line) and feature at $T$ = 80 K (vertical arrow).}
\end{figure}
	
To track the evolution of the quasiparticle and AM dynamics across the thermal CDW transition \TCDW, we investigated the temperature dependence of the transient response in the non-perturbative regime, i.e. for low fluence (85 \uJcm), in the range $T$ = 20 to 150 K. Fig.~\ref{fig:dynamics}\textbf{a} shows the temperature dependent dynamics extracted at 2.14 eV. The magnitude of the incoherent signal (\delR) is representative of the photoexcited quasiparticle population and dependent on the electronic structure \cite{Demsar1999,Torchinsky2013}. In Fig.~\ref{fig:dynamics}\textbf{b}, we track the minimum in the \delR~signal (purple diamonds) which occurs in the first few hundred femtoseconds and corresponds to the magnitude of the negative 2.14 eV feature (see Fig.~\ref{fig:response}\textbf{d}), as discussed previously. Here, we clearly see that this feature emerges below \TCDW~= 110 K as the signal becomes strongly negative, demonstrating that there is change in the quasiparticle dynamics or electronic structure at the expected transition temperature due to the reconstruction induced by the CDW, in full agreement with the interpretation of the electrical resistance in Fig.~\ref{fig:intro}\textbf{a}. Similarly, by analyzing the \delR~signal at 4000 fs (orange squares), i.e. after significant oscillatory dephasing has taken place, we are able to study the dynamics in the absence of strong modulation from the AM. Interestingly however, we find that after the initial change at \TCDW, the 4000 fs signal reaches a local minimum at $T$ = 80 K where it then reverses direction, moving from negative to positive signal at this photon energy. By analyzing various probe photon energies across the measured spectrum (see Supporting Information B), we observe similar signatures at both 110 K and 80 K in the quasiparticle dynamics. 

\begin{figure*}
    \includegraphics[width=1.0\linewidth,clip]{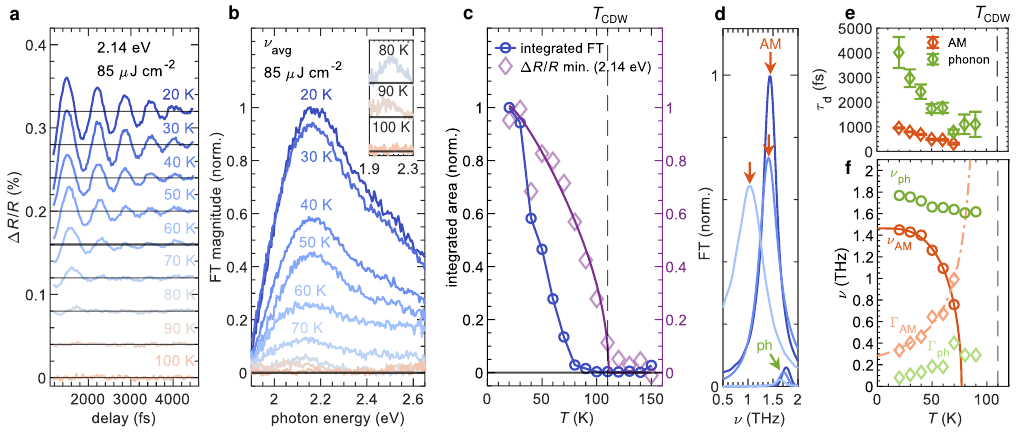}
    \caption{\label{fig:oscillations} \textbf{Frequency analysis of the coherent response across the thermal CDW transition}. (\textbf{a}) Temperature dependence ($T$ = 20 - 100 K) of the isolated oscillatory component at 2.14 eV for low fluence (85 \uJcm). (\textbf{b}) Probe photon energy dependence of the FT magnitude of the total coherent response averaged over the frequency range 0.7 - 2.0 THz. The inset shows a zoom of the data in the range  80 - 100 K. (\textbf{c}) Temperature dependence of the integrated FT spectral area in panel \textbf{b}. As a comparison, the normalized magnitude of the minimum transient signal, \delR~at 2.14 eV is also shown (data from Fig.~\ref{fig:dynamics}\textbf{b}), where the solid purple line is a guide to the eye based on a function for second-order phase transitions with $T_\mathrm{c}$~= 110 K. (\textbf{d}) FT of the individual frequency components at selected temperatures $T$ = 20, 40 and 60 K from the fitting procedure using two damped oscillators applied to the data in panel \textbf{a} (colors matched). (\textbf{e}) Temperature dependence of the damping time, $\tau_{\mathrm{d}}$ associated with the AM and CDW phonon. (\textbf{e}) Temperature dependence of the AM ($\nu_{\mathrm{AM}}$) and CDW phonon ($\nu_{\mathrm{ph}}$) frequencies. The solid orange line is fit to $\nu_{\mathrm{AM}}$ using a function for second-order phase transitions with critical temperature $T_\mathrm{AM}$~= 78 K. The associated damping rate $\Gamma = 1 / \pi \tau_{\mathrm{d}}$ of each each component is also shown. The dashed orange line is a guide to the eye for the temperature evolution of  $\Gamma_{\mathrm{AM}}$.}
\end{figure*}

Having established that there is a change in the quasiparticle dynamics at the expected transition \TCDW~= 110 K, and possible anomaly at 80 K, we next focus on the coherent component of the signal to investigate further. Starting with the response near the oscillatory maximum at 2.14 eV, as reported in Fig.~\ref{fig:oscillations}\textbf{a}, we clearly observe a weakening of the oscillation amplitude and increased damping when approaching \TCDW~from below, as expected. However, the strong modulation seems to disappear at lower temperatures for $T \geq$ 80 K.

Since the CDW oscillations are present over a wide spectral range, as discussed previously (see Fig.~\ref{fig:response}\textbf{c}), it is instructive to analyze the full probe photon energy dependence of the FT magnitude, as shown in Fig.~\ref{fig:oscillations}\textbf{b}. Given the limited frequency resolution of our data, and the intrinsically broad linewidth of the AM, we are unable to reliably extract the spectral profiles of the AM and CDW phonon separately by FT. Therefore, we extract the FT spectrum averaged over a range of frequencies 0.7 - 2.0 THz such that it represents the total coherent response i.e. $\nu_{\mathrm{AM}}$ and $\nu_{\mathrm{ph}}$, although we find comparable results if we select single frequencies within this range. Initially at 20 K, the spectral shape of the FT magnitude is similar to that reported previously in Fig.~\ref{fig:response}\textbf{c}, then for increasing temperature, the magnitude reduces rapidly while its peak redshifts slightly, similar to the spectral shape of the transient reflectivity in Supporting Information A. For $T \geq$ 80 K, the FT amplitude appears to falls to near zero, although as shown in the inset of Fig.~\ref{fig:oscillations}\textbf{b}, there is still weak intensity observable above the noise floor until 100 K. The temperature dependence of the spectrally integrated FT amplitude is reported in Fig.~\ref{fig:oscillations}\textbf{c} (circles, left axis) and compared to the incoherent component of the signal, \delR~min. (diamonds, right axis). The incoherent component clearly exhibits a behaviour expected for a second-order phase transition with \TCDW~= 110 K, with significant onset precisely at \TCDW~which grows rapidly, followed by saturation behaviour towards zero temperature. By contrast, the coherent signal exhibits a behavior that is not typical of such a transition and grows-in for temperatures significantly lower than the expected \TCDW. Indeed, we observe a weak onset of the coherent signal below $\sim$ 100 K until 80 K, where it suddenly undergoes an increase in magnitude by $\sim$ 5 times between 80 and 70 K. Towards lower temperature, the magnitude grows with superlinear behaviour. 

The apparent anomalous behaviour of the coherent response at 80 K matches well with the signatures observed in the incoherent dynamics (see Fig.~\ref{fig:dynamics}\textbf{b}). To investigate further, we analyze the data in Fig.~\ref{fig:oscillations}\textbf{a} using the fitting procedure described previously based on two damped oscillators (see Fig.~\ref{fig:frequency}\textbf{a}), allowing us to extract the frequency and damping time for both the AM and CDW phonon as a function of temperature. Fig.~\ref{fig:oscillations}\textbf{d} shows the FT of the individual frequency components of the fit at selected temperatures to illustrate the overall trend. For increasing temperature, we see that the intensity of the AM reduces rapidly (orange arrow) in addition to a strong redshift and increase in linewidth. By comparison, the CDW phonon (green arrow), exhibits a weaker temperature dependence. Fig.~\ref{fig:oscillations}\textbf{e} and Fig.~\ref{fig:oscillations}\textbf{f} report the full temperature dependence of the parameters up to 70 K using two damped oscillators ($\nu_{\mathrm{AM}}$ and $\nu_{\mathrm{ph}}$), beyond which we cannot reliably fit the data with this model, instead using a single oscillator (with frequency $\nu_{\mathrm{ph}}$ only) for 80 K and 90 K, before the magnitude of the total oscillatory signal approaches the noise floor at 100 K (see inset Fig.~\ref{fig:oscillations}\textbf{b}). In Fig.~\ref{fig:oscillations}\textbf{f}, we confirm the relatively weak temperature dependence of the $\sim$ ~1.8 THz phonon frequency, while the suspected AM initially at $\sim$ 1.45 THz ($\sim$ 48 \cm) at 20 K softens dramatically towards higher temperatures, eventually reaching $\sim$ 0.78 THz ($\sim$ 26 \cm) at 70 K. Hence, we confirm the low frequency mode, initially at $\sim$ 1.4 THz, as the \textit{soft mode} or AM of the CDW \cite{Demsar1999,Demsar2002,Mohr-Vorobeva2011,Torchinsky2013,Werdehausen2018,Sayers2020b,Sayers2022,Sayers2023a,Tuniz2023a}.

Crucially, we find that the temperature dependence of the AM can be fitted (orange solid line in Fig.~\ref{fig:oscillations}\textbf{f}) using a semi-phenomenological model for second-order phase transitions, which has been used to describe the CDW order parameter in TMDs \cite{Chen2015,Chen2016,Chen2018,Feng2018,Chen2022}:
	
\begin{equation}\label{eq:bcs}
	\Delta(T) \propto tanh\left( k \sqrt{\frac{T_{\mathrm{AM}} - T}{T}} \right)
\end{equation}
	
\noindent
We find a prefactor $k$ = 1.82 and a critical temperature of $T_{\mathrm{AM}}$ = (78 $\pm$ 2) K where the AM softens to zero frequency, coinciding precisely with both the onset of the intense coherent response, and signatures in quasiparticle dynamics occurring at 80 K, as discussed previously. 

Based on the result in Fig.~\ref{fig:oscillations}\textbf{e}, we can estimate the damping rate given by $\Gamma = 1 / \pi \tau_{\mathrm{d}}$ and compare it with the mode frequencies. In Fig.~\ref{fig:oscillations}\textbf{f}, we clearly see that there is crossover of the AM frequency and its damping rate at $\sim$80 K, corresponding to the over-damped response when $\nu_{\mathrm{AM}}$ = $\Gamma_{\mathrm{AM}}$ \cite{Zeiger1992}, while the CDW phonon displays no such crossover and is expected to be present up to 110 K based on the trends of $\nu_{\mathrm{ph}}$ and $\Gamma_{\mathrm{ph}}$. This is consistent with the total coherent response in Fig.~\ref{fig:oscillations}\textbf{b} and \textbf{c}, where the weak signal which appears below 100 K likely arises from the CDW phonon alone, before the sharp rise between 80 and 70 K marks the onset of the AM. 

Therefore, based on our analysis, we identify two distinct temperatures of interest in \VSe: (i) An initial second-order change in the electronic structure at the expected \TCDW~= 110 K accompanied by the onset of phonons of the CDW lattice, followed by (ii) a weak anomaly in the quasiparticle dynamics, and the appearance of the AM at $T_{\mathrm{AM}}$ $\approx$ 80 K which grows-in with superlinear behaviour. In principle, a crossover to critical damping at temperatures lower than $T_{\mathrm{c}}$~can occur for any second-order phase transition as a result of the simultaneous mode frequency softening and diverging damping rate towards the critical temperature. Such behaviour can explain the loss of resolvable CDW oscillation amplitude for temperatures below \TCDW~in 2$H$-TaSe$_2$ \cite{Demsar1999,Demsar2002} and 1$T$-TiSe$_2$ \cite{Mohr-Vorobeva2011} for example, although mode softening still tends towards \TCDW, not below. Here instead, we believe that our results provide spectroscopic evidence for a second CDW-related transition in \VSe, given the dramatic softening towards $\sim$80 K and the coincidence of this temperature with previous reports in the literature, which we discuss in the following. 

Previous x-ray diffraction \cite{Tsutsumi1982} and electronic transport \cite{Thompson1979} experiments reported signatures of an incommensurate charge-density-wave (ICCDW) transition occurring at 110 K, followed by a \textit{lock-in} transition to a commensurate charge-density-wave (CCDW) below 85 K involving a small change in the out-of-plane component from $q_z$ = 0.314$c_0$ to 0.307$c_0$ \cite{Tsutsumi1982}. Indeed, the onset of multiple CDW transitions with increasing commensurability as temperature is reduced is a well-known phenomenon in other group-5 TMDs \cite{Wilson1975}. In this scenario, since the charge order is incommensurate with the lattice below 110 K, the strong dephasing leads to an overdamped response of the AM, as noted in other systems \cite{Tsang1977,Demsar2002,Perfetti2008,Sayers2020b}, such that it cannot be well-resolved in the time-domain. Then below 80 K, the charge order locks-in to the lattice periodicity, meaning that the dephasing becomes suddenly weaker, and subsequently the oscillatory amplitude increases. Notable signatures of a CDW-related transition at 80 K were also detected by electron diffraction \cite{Eaglesham1986} experiments in which a 3$q$-2$q$ transition was observed and possibly linked with the lock-in transition \cite{Tsutsumi1982}, which was suggested to be the intrinsic behaviour of true stoichiometric \VSe. Here, the typical second-order 3$q$ CDW first emerges below 110K, which then transforms into a pattern of 2$q$-domains at 80 K accompanied by a sharp increase in the CDW amplitude. This was later supported by scanning tunnelling microscopy experiments \cite{Giambattista1990} which, at the lowest temperature of 4.2 K, found a commensurate CDW but with one of the three $q$-vectors absent or weak, which varied across the sample surface, suggestive of a domain structure. Free energy calculations suggest that the energy difference between the 3$q$ and 2$q$ states is small~\cite{Eaglesham1986}, compatible with the reported transitions temperatures. Based on our observations, it is plausible that the observed AM belongs to the 2$q$ (CCDW) state which softens towards 80 K, while the CDW amplitude of the 3$q$ (ICCDW) state may be intrinsically weak, consistent with the observations of Ref. \cite{Eaglesham1986}.

\subsection{Photoinduced CDW melting in out-of-equilibrium conditions}
	
To provide further information about the quasiparticle and AM dynamics and their contribution to the CDW phase, we explored the out-of-equilibrium dynamics of \VSe~by performing a pump fluence dependence in the range $F$ = 85 to 690 \uJcm~at a fixed temperature ($T$ = 4.3 K) such that the system is initially deep in the CDW phase.
	
\begin{figure*}
    \includegraphics[width=1.0\linewidth,clip]{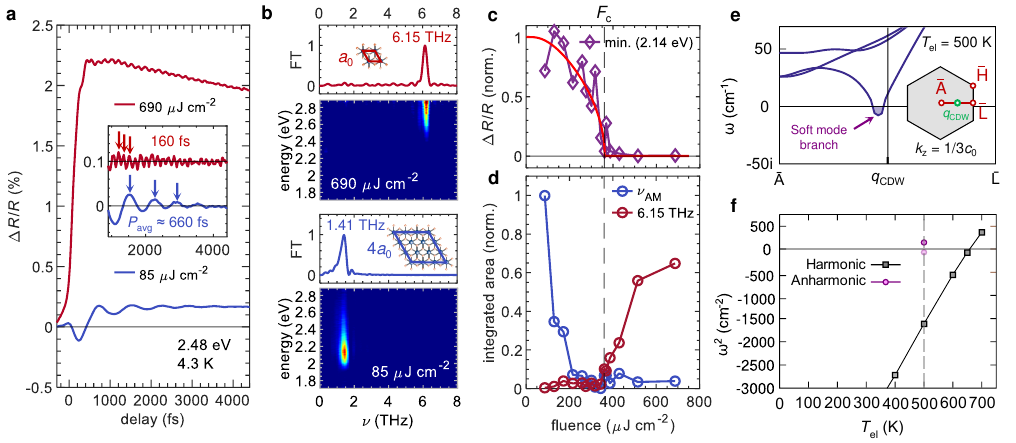}
    \caption{\label{fig:photoinduced} \textbf{Evidence for photoinduced CDW transition in out-of-equilibrium conditions.}. (\textbf{a}) Comparison of the transient response extracted at 2.48 eV for low (85 \uJcm) and high (690 \uJcm) fluence excitation at $T$ = 4.3 K. The inset shows the isolated coherent components in each case, where the arrows highlight maxima of the effective instantaneous periods. The average period, $P_{\mathrm{avg}}$ is $\sim$660 fs and $\sim$160 fs in the low and high fluence regime, respectively. (\textbf{b}) Fourier transform (FT) analysis showing the spectral dependence of the oscillation amplitude in each regime. The coloured traces show the FT amplitude extracted near the oscillatory maximum at 2.14 eV (low fluence) and 2.80 eV (high fluence). (\textbf{c})  Fluence dependence of the absolute magnitude of the minimum in the \delR~signal at 2.14 eV. The red line is a fit with critical fluence $F_{\mathrm{c}}$~= 360 \uJcm. (\textbf{d}) Fluence dependence of the integrated FT spectral area of the AM versus the 6.15 THz mode. (\textbf{e}) DFT calculations of the phonon dispersion along the A-L direction ($k_z$ = 1/3$c_0$) within the harmonic (SSCHA) approximation using an electron and lattice temperature of $T_{\mathrm{el}}$ = 500 K and $T_{\mathrm{ph}}$ = 4 K. The inset illustrates the path through the Brillouin zone. (\textbf{f}) Square of the lowest phonon frequency, $\omega^2$ at \qCDW~as a function of electronic temperature in the harmonic approximation (squares), compared to SSCHA (circles). The light pink circle indicates the absolute lowest eigenvalue, $\omega_{\mathrm{min}}^2$ in the entire Brillouin zone for the SSCHA calculation.}
\end{figure*}

Fig.~\ref{fig:photoinduced}\textbf{a} compares the transient response for low (85 \uJcm) and high (690 \uJcm) fluence regimes where we identify a different behaviour. Aside from the expected larger incoherent signal in \delR~at high fluence, the most significant difference is the change in the dominant oscillatory period from $\sim$ 660 fs to $\sim$ 160 fs, as highlighted by the inset in Fig.~\ref{fig:photoinduced}\textbf{a}. FT analysis in Fig.~\ref{fig:photoinduced}\textbf{b} confirms a switch in the dominant frequency from the AM at $\sim$ 1.41 THz to a different mode at $\sim$ 6.15 THz ($\sim$ 205 \cm). The high frequency mode is in good agreement with the \Ag~mode of the normal phase structure as measured by Raman spectroscopy in Fig.~\ref{fig:intro}\textbf{c} and in previous studies \cite{Sugai1985}, although here it is slightly redshifted. Such behaviour has been observed in time-domain studies of related systems such as 1$T$-TiSe$_2$, and is a signature of the transient suppression of the CDW lattice distortion \cite{Hedayat2019,Hedayat2021}. Hence, we conclude that at 690 \uJcm, the photoexcitation induces a transient change in the crystal lattice symmetry from the distorted CDW phase ($4a \times 4a$) to normal phase ($a\times a$) i.e. a \textit{photoinduced transition}. Additional evidence for a photoinduced transition are obtained from the FT maps in Fig.~\ref{fig:photoinduced}\textbf{b} (see also Supporting Information C), where the spectral profile of the oscillation amplitude is different in the two regimes, moving from lower energies with a maximum near 2.14 eV (corresponding to IB1) for the 1.41 THz AM to higher energies towards 2.80 eV (corresponding to IB2) for the 6.15 THz normal phase mode, indicating a rearrangement of the electronic structure, and thus optical transitions relevant to the dominant mode in each case \cite{Sayers2022}.

In a photoinduced transition scenario, it is expected that the increased photocarrier density, or equivalently enhanced electronic temperature \Tel, results in unfavourable conditions for the CDW to remain stable, while the lattice temperature \Tph, remains effectively cold (\Tph $<$ \TCDW) and is therefore referred to as a non-thermal phase transition \cite{Mohr-Vorobeva2011}. Performing experiments as a function of the excitation density then allows determination of the critical fluence, \Fc~required to melt the CDW state, giving an indication of its strength. Hence, we next focus on the results of the fluence dependence in the range $F$ = 85 to 690 \uJcm~in order to track the dynamics across the photoinduced phase transition , identify the critical fluence \Fc, and compare the behaviour to the thermal transition in the non-perturbative regime.

From the transient spectra, we find that the 2.14 eV feature, previously ascribed to the CDW phase, weakens and eventually disappears for increasing fluence between $\sim$ 340 and 390 \uJcm~(see Supporting Information D). In Fig.~\ref{fig:photoinduced}\textbf{c}, we plot the minimum in the \delR~signal at 2.14 eV as a function of fluence, and fit the data with Equation~\ref{eq:bcs} which allows us to determine a critical fluence of \Fc~= (360 $\pm$ 5) \uJcm~which we assign to the photoinduced CDW-to-normal phase transition. It should be noted that this threshold is relatively high considering the moderate transition temperature (110 K) and weak order parameter ($\Delta$ = 10 meV) of \VSe. This is at odds with what would be expected for a purely FSN scenario since any small perturbation of the ideal nesting conditions \cite{JohannesMazin2008}, such as a shift in chemical potential or a broadening of states near $E_\mathrm{F}$, e.g. due to an increase in electronic temperature, should destroy the CDW. In systems with weak electronic order such as 2$H$-NbSe$_2$, the threshold can be very low \Fc~$\approx$ 10 \uJcm~\cite{Venturini2023}, while in systems such as 1$T$-TiSe$_2$ where an excitonic mechanism is thought to contribute, significant suppression is found at low-moderate fluences ($\approx$ 40 - 200 \uJcm) due to Coulomb screening effects ~\cite{Mohr-Vorobeva2011,Porer2014,Hedayat2019}. Instead, in systems where the lattice order has been suggested to play a dominant role, the CDW phase is notably robust, for example in the Kagome metal ScV$_6$Sn$_6$ (\Fc~$\approx$ 1000 \uJcm) \cite{Tuniz2023b}, or 1$T$-TaSe$_2$ where the CDW cannot be melted even for fluences up to 3000 \uJcm~\cite{Sayers2020b,Sayers2023a}. Similar to other quasi-2D systems, the CDW transition in \VSe~is more likely to depend on a momentum-dependent EPC mechanism \cite{Rossnagel2011,Zhu2015,Henke2020}, which is consistent with our observations of the moderate fluence required to melt the CDW.

Next, we compare the threshold fluence determined from the incoherent electron dynamics with the behaviour of the coherent response across the photoinduced transition. In Fig.~\ref{fig:photoinduced}\textbf{d}, we plot the integrated spectral area of the FT magnitude (same procedure as Fig.~\ref{fig:oscillations}\textbf{b}-\textbf{c}), extracted at $\nu_{\mathrm{AM}}$ and 6.15 THz, as a function of fluence. Firstly considering the progression from low to high fluence, we observe that the 6.15 THz mode, indicative of the normal phase structure, has near zero intensity before appearing above \Fc~= 360 \uJcm, in excellent agreement with the threshold determined previously from the quasiparticle dynamics, confirming a simultaneous reconstruction of the electronic and lattice structure at this fluence. By comparison, considering the progression from high to low fluence, the AM at 1.41 THz does not appear strongly below this critical fluence but instead only becomes well-resolved below $\sim$ 235 \uJcm. Therefore, similar to the thermal transition, we determine two distinct fluences corresponding to (i) the initial reconstruction of the electronic and lattice structure at $F_{\mathrm{c}}$~= 360 \uJcm, followed by (ii) the subsequent onset of the AM below $F_{\mathrm{AM}}$~= 235 \uJcm. Comparing the ratios for the thermal ($T_{\mathrm{CDW}}$/$T_{\mathrm{AM}}$) and photoinduced ($F_{\mathrm{c}}$/$F_{\mathrm{AM}}$) transitions reveals that the AM onset is $\sim 70 \%$ lower than the upper threshold for the normal-to-CDW transition in both cases. Hence, the observation of two distinct critical fluences related to the photoinduced transition in \VSe~corroborates our findings for the thermal phase transition obtained in the non-perturbative excitation regime, although additional information is gained from the former by tracking the onset of the oscillatory signal belonging to the normal phase structure.

\subsection{First-principles simulations of the photoinduced CDW melting}
	
To gain further insights into the photoinduced CDW transition in \VSe, we performed first-principles calculations of the phonon dispersion using density functional theory (DFT), including finite temperature effects via a Fermi-Dirac smearing of the electronic occupation \cite{Baroni2001}. A description of the computational details is included in the Methods section. A CDW instability is indicated by the presence of an imaginary phonon eigenvalue in correspondence to the expected CDW wavevector, \qCDW~\cite{Bianco2017,Monacelli2021}. By introducing an electronic temperature $T_{\mathrm{el}}$, the effective CDW transition temperature can then be determined by the lowest temperature at which the phonon frequency at $q$ = \qCDW~becomes non-imaginary.
	
Following this approach, we first calculated the phonon dispersion within the \textit{harmonic} approximation as a function of temperature, where the electronic (\Tel) and lattice (\Tph) subsystems are in thermal equilibrium, i.e. \Tel~=~\Tph. At the lowest temperatures, we find negative phonon frequencies along the \ABar-\LBar~direction in the vicinity of \qCDW~indicating an instability (see Supporting Information F). In Fig.~\ref{fig:photoinduced}\textbf{f}, we investigate the evolution of this instability by plotting the square of the lowest phonon frequency, $\omega^2$ found at \qCDW~as a function of the electronic temperature. Above \Tel~= 650 K, we find that $\omega^2$ becomes positive indicating that the instability has disappeared and thus we define this as the predicted CDW transition temperature. Hence, we have demonstrated that it is possible to melt the CDW phase at elevated electron temperatures, although clearly the \textit{harmonic} approach overestimates the experimental transition temperature, which has also been noted in previous theoretical investigations \cite{Diego2021,Falke2021}.
	
Alternatively, it has been shown that by including the effects of anharmonicity, it is possible to achieve more realistic estimates in some systems, including \VSe~\cite{Diego2021}. Therefore, we also calculated the phonon dispersion including \textit{anharmonic} effects using the stochastic self-consistent harmonic approximation (SSCHA). In this case, we can define separate temperatures for the electronic and lattice subsystems, which is particularly useful for the out-of-equilibrium situation in ultrafast spectroscopy experiments where the dominant effect of photoexcitation is to induce a transiently enhanced electronic temperature \cite{Giannetti2016,Hedayat2019}, while the lattice remains effectively cold within the first few picoseconds. For the purposes of the simulation, we chose \Fc~= 350 \uJcm~for the critical fluence and estimated the corresponding critical photocarrier density to be $n_{c} \approx 1.50 \times 10^{20}$ cm$^{-3}$ based on the experimental conditions used in this work and the optical properties of our \VSe~samples determined previously \cite{Feng2021} (see Supporting Information E). From this, we obtain an initial maximum electronic temperature $T_{\mathrm{el, max}}$ = 780 K. As a conservative estimate for the anharmonic SSCHA calculations, we use \Tel~= 500 K and lattice temperature \Tph~= 4 K to simulate the out-of-equilibrium conditions (see Supporting Information F). The results of the SSCHA phonon dispersion are shown in Fig.~\ref{fig:photoinduced}\textbf{e} along the \ABar-\LBar~direction, as illustrated by the inset. We find a positive phonon frequency at \qCDW~suggesting a melting of the CDW phase, while a weak instability remains ($\omega \leq 10i$ cm$^{-1}$) slightly away from \qCDW~towards the \ABar-point, highlighted by the shaded area. These results are also reported in Fig.~\ref{fig:photoinduced}\textbf{f} where the circles show the value of $\omega^2$ at \qCDW~(dark pink) and the minimum value found across the entire BZ (light pink). Hence, the SSCHA results indicate that it is possible to induce CDW suppression for transiently elevated electronic temperatures, and that the out-of-equilibrium conditions (\Tel~= 500 K, \Tph~= 4 K) based on the experimentally determined critical fluence \Fc, are very close to the CDW-to-normal phase transition threshold. Thus, the simulations provide further support for the moderate excitation densities required for CDW melting in \VSe, consistent with a dominant EPC mechanism, for example via photoinduced screening of the electron-phonon interaction by hot carriers.


\section*{Conclusion}

We studied the quasiparticle and AM dynamics of \VSe~using femtosecond broadband transient reflectivity experiments. Despite detecting clear signatures in the temperature-dependent quasiparticle dynamics of a rearrangement of the electronic structure at the expected CDW transition \TCDW~= 110 K, we found that the AM instead exhibits anomalous behaviour as it softens to zero frequency at a lower temperature of $T_{\mathrm{AM}}$ $\approx$ 80 K. Our results provide spectroscopic evidence for the presence of an additional CDW-related transition occurring at 80 K in \VSe~which may be related to incommensurability (ICCDW-to-CCDW) or domain formation (3$q$-to-2$q$ transition), and deserves further experimental study. In addition, we demonstrated a photoinduced melting of the CDW phase at moderate excitation densities, supported by first-principles calculations, which is inconsistent with a purely Fermi surface nesting mechanism for the CDW. Instead we suggest that a dominant electron-phonon coupling mechanism is more likely to be significant, similar to other TMDs.
	
	
\section*{Methods}
	
\subsection*{Crystal growth \& characterization}
	
Single crystals of \VSe~were grown by chemical vapour transport (CVT) at 550$^\circ$C with iodine as the transport agent, as detailed in Ref. \cite{Sayers2020a}. Electrical transport measurements confirmed a normal-to-CDW phase transition at \TCDW ~= 112 K and residual resistance ratio $RRR$ $\approx$ 50, indicative of the high sample quality. A single crystal of $\sim$ 3 mm$^2$ was mounted inside a liquid helium flow cryostat and cleaved prior to experiments to expose a pristine surface.

\subsection*{Ultrafast optical spectroscopy}
	
Time-resolved reflectivity experiments were performed using an amplified Ti:sapphire laser (Coherent Libra) which outputs pulses (800 nm, 100 fs) at 2 kHz repetition rate. A portion of the 800 nm ($\sim$1.55 eV) laser fundamental was used as the pump beam, while the remainder was used to produce the broadband probe beam via white light continuum generation in sapphire. The pump beam was modulated at 1 kHz using a mechanical chopper, and the differential reflectivity (\delR) was measured as a function of pump-probe delay time using a mechanical delay line. Pump and probe beams were focused onto the sample at near-normal incidence. Cross polarization was used to avoid scattering artefacts. A portion of the reflected probe in the range 430 - 725 nm ($\sim$1.7 - 2.9 eV) was selected for detection using a spectrometer with silicon CCD array.
	
\subsection*{Computational details}
	
First-principles calculations have been performed within the density functional theory (DFT) framework, as implemented in the Quantum ESPRESSO package \cite{Giannozzi2009,Giannozzi2020}. A kinetic energy cutoff of 40 Ry was used for the plane-wave expansion of the Kohn-Sham wavefunctions and 450 Ry for the density. For the exchange-correlation potential, the generalized gradient approximation in the Perdew, Burke and Ernzerhof (PBE) \cite{Perdew1996} formulation was used. A 24$\times$24$\times$16 Monkhorst-Pack wave-vector grid \cite{Monkhorst1976} was used for the integration of the Brillouin zone. Long-range dipolar van der Waals corrections were included following the approach of Ref.~\cite{Grimme2006}. Experimental lattice parameters were taken from Ref.~\cite{Diego2021} and the internal positions were relaxed, not fixed by symmetry, to internal forces of the order of 10$^{-6}$ Ry/a.u. or lower. Finite temperature effects have been simulated using a Fermi-Dirac distribution.
	
The anharmonic phonon frequencies were calculated within the stochastic self-consistent harmonic approximation (SSCHA), according to the procedure described in Ref.~\cite{Monacelli2021}. SSCHA requires calculations of forces in a supercell with ionic coordinates displaced from equilibrium following a temperature-dependent Gaussian probability (lattice temperature, \Tph). A 4$\times$4$\times$3 supercell with 3$\times$3$\times$3 electronic momentum grid was used for Brillouin zone integration. This approach has been validated in a previous work on \VSe~in Ref.~\cite{Diego2021}. In order to simulate the out-of-equilibrium conditions in the experiment, an elevated electronic temperature of \Tel~= 500 K was used, while the lattice temperature was fixed at \Tph~= 4 K.
	
	
\section*{Acknowledgements}
	
The authors acknowledge the contribution of Liam Farrar and Simon J. Bending for prior electrical characterization of the \VSe~samples. During submission of the manuscript, the authors became aware of a recent study by Sutar \textit{et al.}~\cite{Sutar2024} which corroborates many of the findings presented in this work. This project has received funding from the European Union's Horizon 2020 research and innovation programme Graphene Flagship under grant agreement No. 881603, and ERC grant DELIGHT No. 101052708. The authors acknowledge support from the PRACE program, including access to Joliot-Curie at GENCI@CEA, France, project number 2021240020. GC acknowledges support from European Union’s NextGenerationEU Programme with the I-PHOQS Infrastructure [IR0000016, ID D2B8D520, CUP B53C22001750006] “Integrated infrastructure initiative in Photonic and Quantum Sciences”.

\subsection*{Conflict of Interest:}
	
The authors declare no conflict of interest.

	
\bibliography{refs}




\section*{Supplementary material}

*Supporting Information for*:

# Anomalous amplitude mode dynamics below the expected charge-density-wave transition in $1T$-VSe$_2$

Charles J. Sayers,[1] Giovanni Marini,[2,3] Matteo Calandra,[2,3] Hamoon Hedayat,[4] Xuanbo Feng,[5] Erik van Heumen,[5] Christoph Gadermaier,[1] Stefano Dal Conte,[1] and Giulio Cerullo[1]

[1]*Dipartimento di Fisica, Politecnico di Milano, 20133 Milano, Italy*

[2]*Department of Physics, University of Trento,*
*Via Sommarive 14, 38123 Povo, Italy*

[3]*Graphene Labs, Fondazione Istituto Italiano di Tecnologia,*
*Via Morego, I-16163 Genova, Italy*

[4]*University of Cologne, Institute of Physics II, 50937 Cologne, Germany*

[5]*Institute of Physics, University of Amsterdam,*
*Science park 904, 1098 XH Amsterdam, The Netherlands*

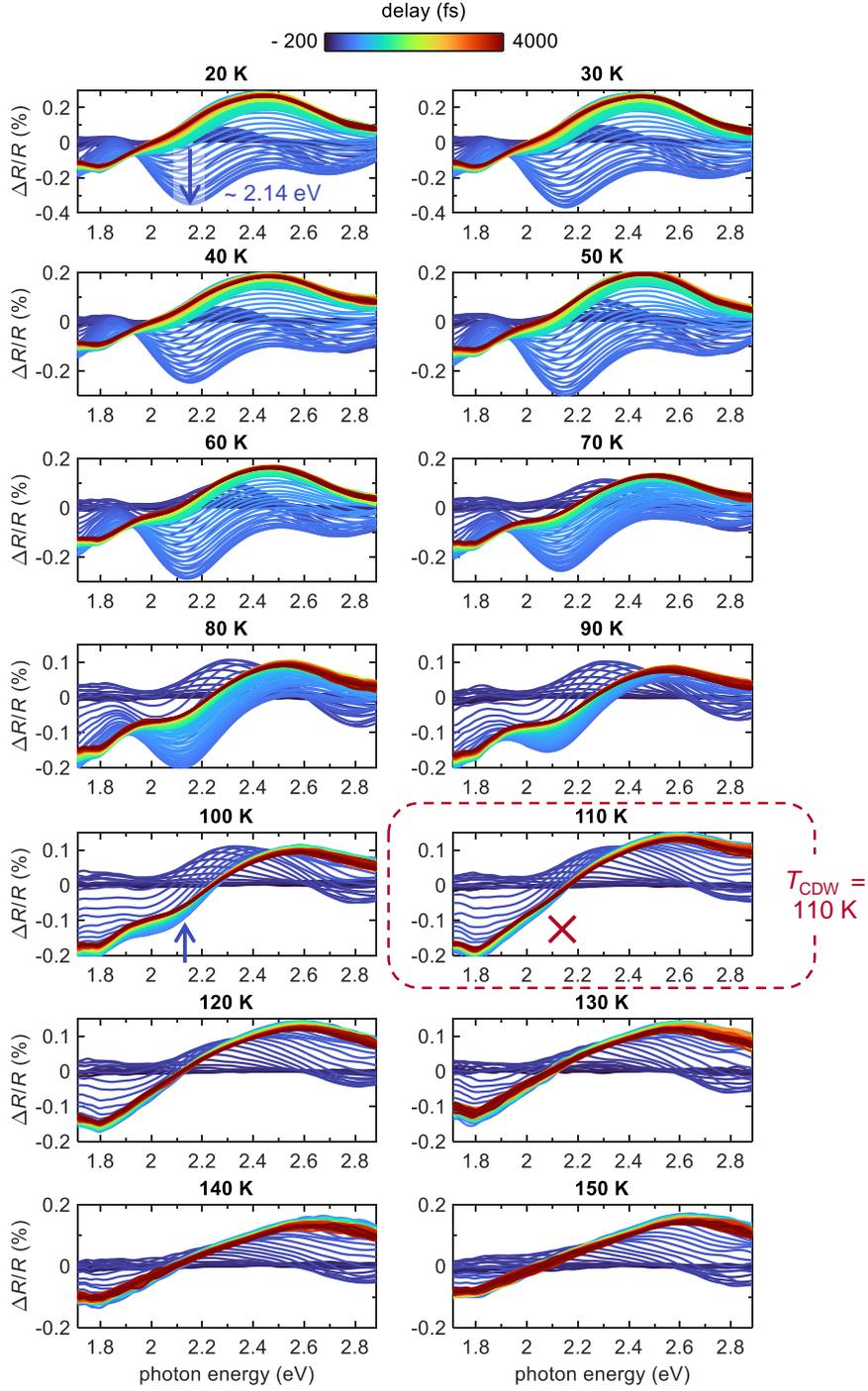

FIG. S1. **Temperature dependence of the transient spectra**. Each panel shows the temporal evolution of the differential spectra, $\Delta R/R$ for temperatures in the range $T = (20 - 150)$ K, as indicated. The color (see color bar) corresponds to the measured pump-probe delay time $t = -200$ to 4000 fs with steps of 20 fs. The expected CDW transition at $T_{\text{CDW}} = 110$ K is indicated.

## A. TEMPERATURE DEPENDENCE OF THE TRANSIENT SPECTRA

Fig. S1 shows the temporal evolution of the differential spectra, $\Delta R/R$ for each temperature in the range $T = (20 - 150)$ K for low pump fluence (85 µJ cm$^{-2}$). As discussed in the main text, the negative feature (initially at $\sim 2.14$ eV for $T = 20$ K) is found for all temperatures $T < T_{\text{CDW}}$ (blue arrow), whereas it is absent for $T \geq T_{\text{CDW}}$ (red cross). Hence, we assume that this feature is closely linked to the CDW, such as an optical transition that becomes allowed or enhanced in the reconstructed phase, for example due to the appearance of a folded band or modification of the density of states. The 2.14 eV feature is particularly evident at early time delays as shown by the spectra with blue colour (see color bar) for all temperatures $T \leq 100$ K, but also remains at long delays up to 4000 fs as seen in the spectra with red colour.

## B. SIGNATURES OF THE CDW TRANSITION IN THE QUASIPARTICLE DYNAMICS AT VARIOUS PHOTON ENERGIES

Fig. S2 shows temperature-dependent analysis of the quasiparticle dynamics at various probe photon energies for low pump fluence (85 µJ cm$^{-2}$). In the main text, analysis of the dynamics was performed at 2.14 eV on the high-energy side of IB1 (see Fig. S2**a**), near to the oscillatory maximum of the AM and is therefore naturally convoluted with the strong oscillatory component, particularly at early time delays. Fig. S2**b** instead shows dynamics extracted at 1.90 eV, which was determined to be near the oscillatory minimum of the AM (see main text Fig 2**c)**, and hence is more representative of the pure electron dynamics in the absence of strong modulation from the AM. Indeed, the temperature dependence of the transient signal at 1.90 eV in Fig. S2**c** confirms a rearrangement of the electronic structure occurring at $T_{\text{CDW}} = 110$ K corresponding to the expected CDW transition temperature. Signatures at 80 K in the dynamics extracted at 4000 fs are either absent or weak. Figs. S2 **d-e** and Figs. S2 **f-g** show analysis of the dynamics at photon energies on the low- (2.48 eV) and high-energy (2.70 eV) side of IB2 (see Fig. S2**a**). Here, we observe signatures at both $T_{\text{CDW}} = 110$ K and $T = 80$ K, supporting our conclusions in the main text. Similar to the analysis at 2.14 eV, we emphasize that signatures at 80 K are detected in the dynamics even at 4000 fs after significant AM dephasing has occurred.

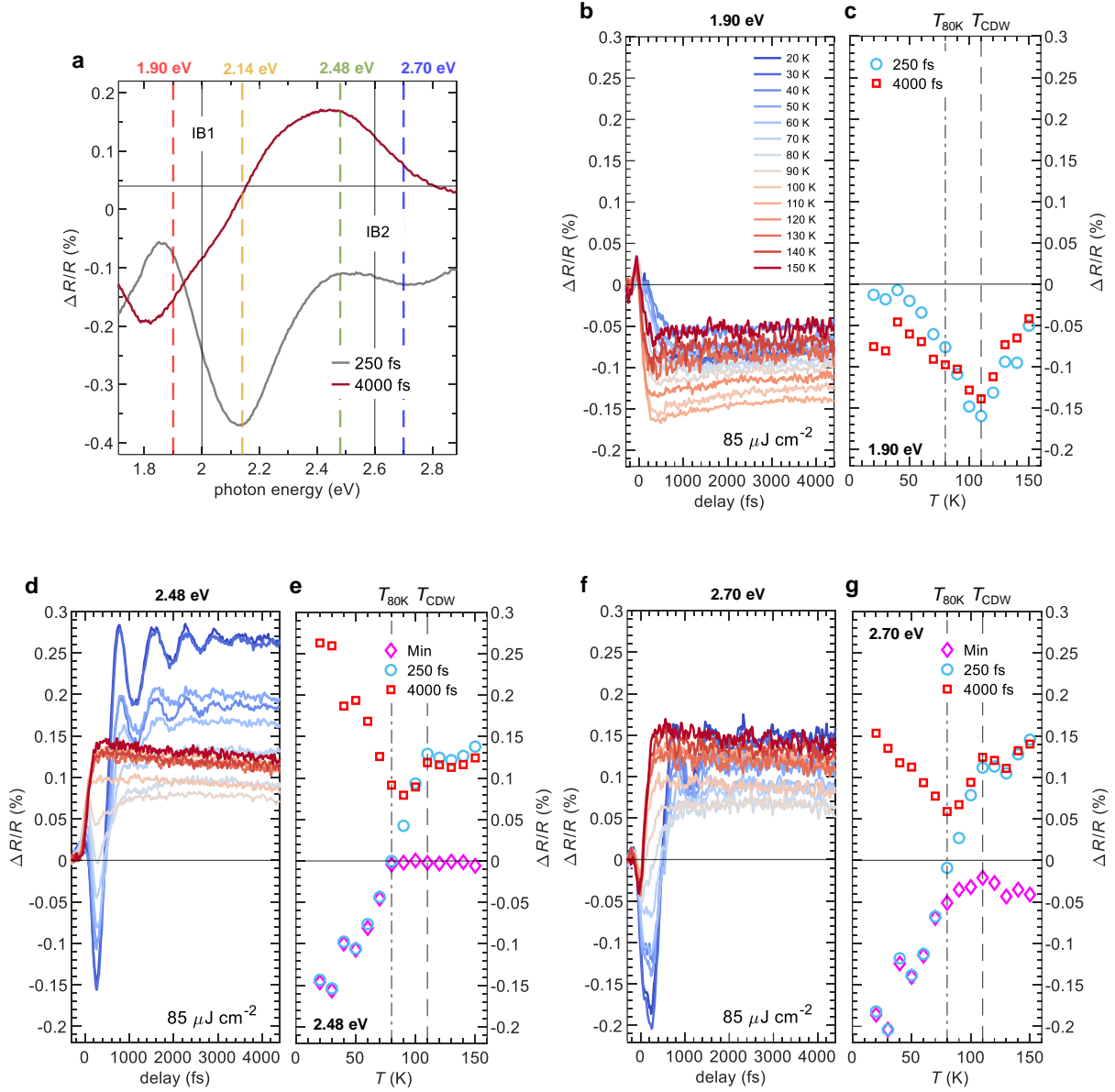

FIG. S2. **Temperature-dependent quasiparticle dynamics at various photon energies**. (**a**) Differential spectra, $\Delta R/R$ at selected time delays, as indicated. The solid vertical lines represent interband transitions IB1 (2.0 eV) and IB2 (2.6 eV) from Ref. [1], while the dashed vertical lines indicate the photon energies for which analysis of the quasiparticle dynamics have been performed (2.14 eV in main text). (**b**) Temperature dependent dynamics extracted at 1.90 eV probe photon energy. (**c**) Magnitude of the transient signal, $\Delta R/R$ at 250 fs and 4000 fs as a function of temperature. Panels (**d**) - (**e**) show the same analysis for 2.48 eV, and similarly panels (**f**) - (**g**) for 2.70 eV. Markers labelled "Min." correspond to the minimum $\Delta R/R$ signal found within a range of -250 to + 1000 fs. Vertical dashed lines indicate $T_{\text{CDW}} = 110$ K and $T = 80$ K.

4

## C. COMPARISON OF THE TRANSIENT RESPONSE WITH INTERBAND TRANSITIONS

Fig. S3 shows the transient response in both the time and frequency domain for the low (85 µJ cm$^{-2}$) and high (690 µJ cm$^{-2}$) fluence regimes, as determined in the main text, and compares the observed features with interband transitions IB1 (2.0 eV) and IB2 (2.6 eV) from Ref. [1]. In the differential reflectivity $\Delta R/R$ maps in Figs. S3 **a** and **b**, we can identify a change of sign and derivative-like shapes corresponding to IB1 and IB2 in both regimes. Notably however, the negative feature at 2.14 eV on the high-energy side of IB1 only appears strongly in the low fluence regime (CDW phase) and corresponds well with the peak in the AM magnitude. By comparison, a weaker negative feature is evident at early times on the high-energy side of IB2 towards 2.80 eV in the high fluence regime (photoinduced normal phase) and this instead corresponds with the intensity profile of the 6.15 THz mode.

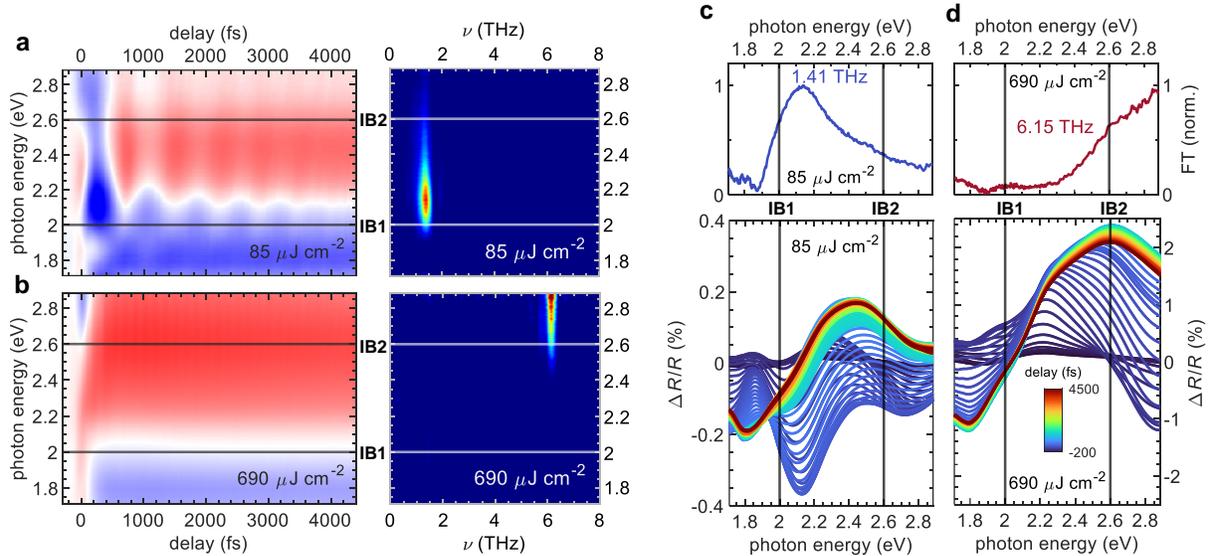

FIG. S3. **Comparison of the transient response with interband transitions**. Differential reflectivity, $\Delta R/R$ maps (left) and FT maps (right) at $T = 4.3$ K for the (**a**) low fluence (85 µJ cm$^{-2}$) and (**b**) high fluence (690 µJ cm$^{-2}$) regimes. Probe photon energy dependence of the normalized FT magnitude extracted at the labelled frequencies (top) and transient spectra (bottom) for the (**c**) low fluence (85 µJ cm$^{-2}$) and (**d**) high fluence (690 µJ cm$^{-2}$) regimes. In all panels, the interband transitions IB1 (2.0 eV) and IB2 (2.6 eV) from Ref. [1] are represented by the horizontal or vertical solid lines, as labelled.

5

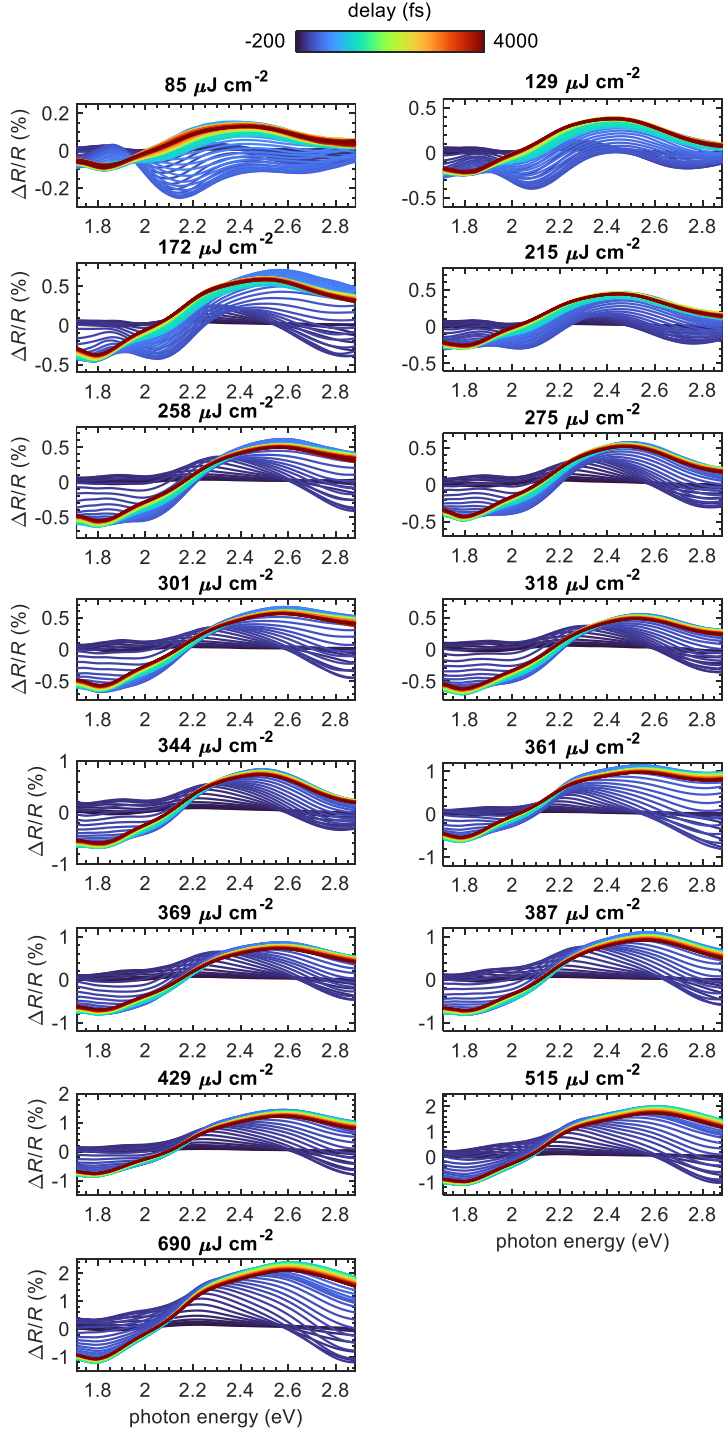

FIG. S4. **Fluence dependence of the transient spectra**. Each panel shows the temporal evolution of the differential spectra, $\Delta R/R$ for pump fluences in the range $F = (85 - 690)$ µJ cm$^{-2}$, as indicated. The color (see color bar) corresponds to the measured pump-probe delay time $t =$ -200 to 4000 fs with steps of 20 fs.

## D. FLUENCE DEPENDENCE OF THE TRANSIENT SPECTRA

Fig. S4 shows the evolution of the differential spectra, $\Delta R/R$ for each fluence in the range $F = (85 - 690)$ µJ cm$^{-2}$ with fixed temperature $T = 4.3$ K. The negative feature previously linked to the CDW phase, and initially at $\sim 2.14$ eV, becomes weak or absent between 344 and 387 µJ cm$^{-2}$. In the main text, the critical fluence, i.e. the fluence required to induce the CDW-to-normal state transition, was determined precisely to be $F_c = (360 \pm 5)$ µJ cm$^{-2}$. This critical fluence is used to inform choices for the input to the first-principles calculations in the following sections.

## E. ESTIMATION OF THE CRITICAL PHOTOCARRIER DENSITY

The initial photocarrier density generated by the pump, $n_0$ for a bulk sample can be estimated assuming a cylindrical excitation volume, and that each single photon generates a single photocarrier, hence:

$$n_0 = \frac{(1-R)(1/\delta)F}{h\nu} \quad (1)$$

where $F$ is the pump fluence and $h\nu \approx 1.55$ eV is the photon energy. Previous experimental data [1] obtained by steady-state optical spectroscopy on the same $1T$-VSe$_2$ samples studied in this work are shown in Fig. S5a and Fig. S5b. This gives a precise value for the absolute reflectivity ($R = 0.48$) and penetration depth ($\delta = 49.4$ nm) at $h\nu \approx 1.55$ eV for $T = 16$ K. The pump fluence is given by $F = P/rA$, where $P$ is the measured laser power, $r = 1$ kHz is the repetition rate and $A = \pi(d/2)^2$ is the pump area incident on the sample based on a Gaussian beam profile with a measured diameter of $d = 385$ µm.

Using Equation 1, we can calculate the photocarrier density for the laser fluence range used in our experiment, $F = (85 - 690)$ µJ cm$^{-2}$ which corresponds to $n_0 = (0.36 - 2.91) \times 10^{20}$ cm$^{-3}$, as shown in Fig. S5c. For the purposes of the first-principles calculations (Section **F**), we chose a critical fluence of $F_c = 350$ µJ cm$^{-2}$, for which we find the corresponding critical photocarrier density to be $n_c = 1.48 \times 10^{20}$ cm$^{-3}$.

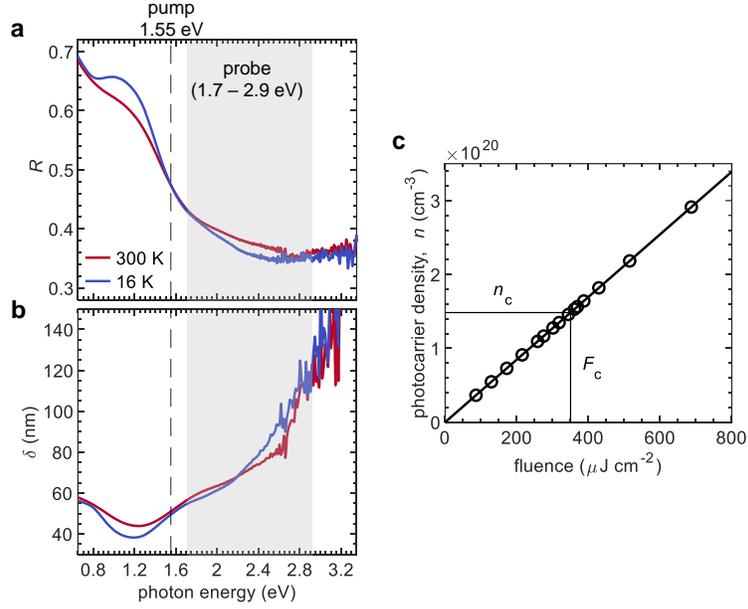

FIG. S5. **Estimation of the photocarrier density**. Steady-state optical spectroscopy data for bulk $1T$-VSe$_2$ from Ref. [1] showing (**a**) the absolute reflectivity, $R$ and (**b**) the penetration depth, $\delta$ obtained by fitting a Drude-Lorentz model to the experimental data. (**c**) Photocarrier density, $n_0$ as a function of pump laser fluence, $F$. The critical photocarrier density, $n_c = 1.48 \times 10^{20}$ cm$^{-3}$ and fluence, $F_c = 350$ µJ cm$^{-2}$ are marked by the horizontal and vertical lines, respectively.

**Error estimations**

Considering the formula for the pump fluence $F = P/rA$, the relative error can be estimated by $\Delta F/F \approx \sqrt{(\Delta P/P)^2 + (\Delta A/A)^2}$. Assuming a maximum percentage error of $\Delta P/P \approx 5\%$ and $\Delta A/A \approx 10\%$ for the measured laser power and pump area respectively, we obtain a value of $\Delta F/F \approx 11\%$. Then considering Equation 1, and the fact that the optical properties of our $1T$-VSe$_2$ samples are extracted from experiment with negligible uncertainty ($\leq 0.1\%$) [1], the error in the photocarrier density becomes proportional to the error in the fluence, $\Delta n_0/n_0 \propto \Delta F/F$. Hence we find the critical photocarrier density with associated error to be $n_c \approx (1.48 \pm 0.16) \times 10^{20}$ cm$^{-3}$ based on $F_c \approx (350 \pm 39)$ µJ cm$^{-2}$.

8

## F. FIRST-PRINCIPLES CALCULATIONS OF THE PHONON DISPERSION: OUT-OF-EQUILIBRIUM APPROACH

To gain further insights into the thermal and photoinduced CDW transition in $1T$-VSe$_2$, we performed first-principle calculations of the phonon dispersion using density functional theory (DFT) including finite temperature effects. A full description of the computational details is given in the Methods section of the main text. When analyzing the phonon dispersion, a CDW instability is signalled by the presence of an imaginary phonon eigenvalue in correspondence to the expected CDW wavevector, $q_{\text{CDW}}$ indicating that the particular mode is unstable there [2, 3]. By applying a Fermi-Dirac smearing of the electronic occupation, we introduce a finite temperature in the calculation [4] which allows us to investigate the effective temperature-dependence of the instability. A meaningful definition of the CDW transition temperature can then be given as the lowest temperature at which the phonon frequency at $q = q_{\text{CDW}}$ becomes non-imaginary.

### Treatment of the photoexcitation process

Photoexcitation of solids using ultrashort laser pulses [5, 6] leads to an out-of-equilibrium situation characterized by a transient increase of the electronic temperature $T_{\text{el}}$, which can be treated using the Fermi-Dirac distribution. In order to estimate the maximum possible electronic temperature, $T_{\text{el,max}}$, we require that the energy transferred from the laser to the sample is completely absorbed by the electrons, disregarding the heat transfer to the lattice, and hence

$$h\nu \times n_{\text{el}} = \Delta U(T_{\text{el,max}}) \tag{2}$$

where $n_{\text{el}}$ is the number of photoexcited electrons and $h\nu = 1.55$ eV is the experimental photon energy. $\Delta U(T_{\text{el,max}})$ is the variation of the electron gas' internal energy which can be expressed as

$$\Delta U(T_{\text{el,max}}) = \int_{-\infty}^{+\infty} d\epsilon \, [u(\epsilon, T_{\text{el,max}}) - u(\epsilon, T_0)] \tag{3}$$

where $u(\epsilon, T)$ represents the electrons' energy density at energy $\epsilon$ and temperature $T$ and can be expressed as a function of the Kohn-Sham eigenvalues as

$$u(\epsilon, \mathrm{T}) = \frac{1}{N_k} \sum_{i,\mathbf{k}} \epsilon_{i,\mathbf{k}}\, f(\epsilon_{i,\mathbf{k}}, T)\, \delta(\epsilon - \epsilon_{i,\mathbf{k}}) \tag{4}$$

where $f(\epsilon_{i,\mathbf{k}}, T)$ is the Fermi-Dirac distribution at temperature $T$, $\delta(\epsilon - \epsilon_{i,\mathbf{k}})$ is the Dirac delta, while the $\mathbf{k}$-sum is extended over the first BZ and $i$ is the band index. In this case, $T_0 = 4$ K is the initial experimental temperature.

Based on the experimental critical photocarrier density $n_c \approx 1.50 \times 10^{20}$ cm$^{-3}$ (see Section **E**), we estimate an initial maximum electronic temperature $T_{\mathrm{el,max}} = 780$ K. For the SSCHA calculations, a conservative value of $T_{\mathrm{el}} = 500$ K was chosen, in order to account for imperfect energy transfer from the laser to the electron subsystem and the fact that the lattice also shares a portion of the specific heat.

**Results of the calculation within the harmonic approximation**

Fig. S6**a** shows the three-dimensional Brillouin zone (BZ) of $1T$-VSe$_2$, while the zoom shows a section at $k_z = 1/3c_0$ which is the expected out-of-plane component of the CDW wavevector [7]. Since it is instructive to analyze the phonon dispersion within this plane, we select the path $\bar{\mathrm{A}}$-$\bar{\mathrm{L}}$-$\bar{\mathrm{H}}$ as indicated by the red line in Fig. S6**a**, which passes through in-plane CDW wavevector, $q_{\mathrm{CDW}} = 1/4a_0$.

We start with calculations of the phonon dispersion within the *harmonic* approximation as a function of temperature by introducing a Fermi-Dirac distribution for the electronic occupation [4]. The system is defined by a single temperature where the electronic and lattice subsystems are in thermal equilibrium, i.e. $T_{\mathrm{harm}} = T_{\mathrm{el}} = T_{\mathrm{ph}}$. The results, shown in Fig. S6**b** at $T_{\mathrm{el}} = 500$ K, clearly demonstrate the presence of negative phonon frequencies along the $\bar{\mathrm{A}}$-$\bar{\mathrm{L}}$ direction in the vicinity of $q_{\mathrm{CDW}}$, indicating an instability. We investigate the evolution of this instability by plotting the square of the lowest phonon frequency, $\omega^2$ found at $q_{\mathrm{CDW}}$ as a function of the electronic temperature as shown in Fig. S6**d**. We observe a reducing instability for increasing temperature, while the linear behaviour of $\omega^2$ with $T_{\mathrm{el}}$ is expected in the Landau description for a second-order phase transition as $\omega \propto (T - T_c)^{1/2}$. Above $T_{\mathrm{el}} = 650$ K, we find that $\omega^2$ becomes positive indicating that the instability has disappeared, and hence we can define this as the predicted CDW transition temperature. This is notably higher than the typically reported experimental value of $T_{\mathrm{CDW}} = 110$ K. Thus, we conclude that the harmonic approximation overestimates the transition temperature, which has also

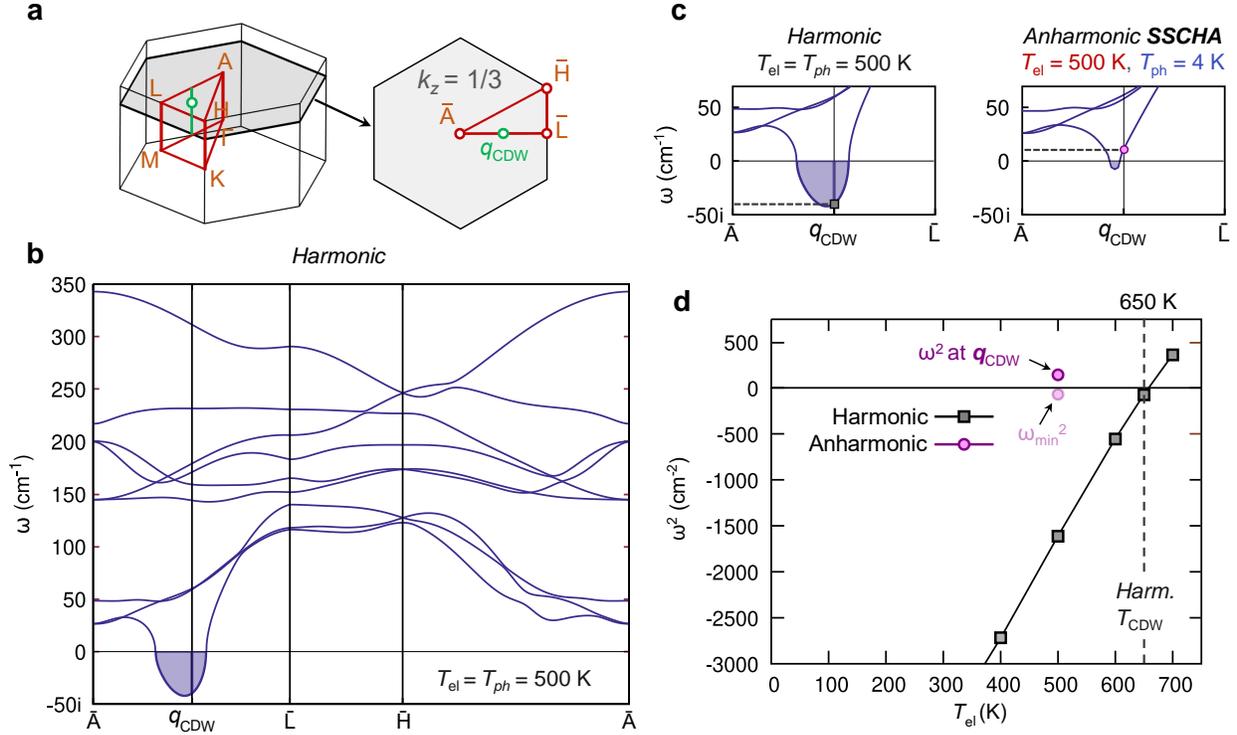

FIG. S6. **Out-of-equilibrium melting of the CDW from first-principles calculations**. (**a**) Sketch of the three-dimensional Brillouin zone of $1T$-VSe$_2$. The shaded hexagon represents a two-dimensional (2D) section at $k_z = 1/3 c_0$ close to the expected CDW wavevector, $q_{\text{CDW}}$ in the $z$-dimension. Projected high-symmetry points onto the 2D plane are labelled in orange. The green marker represents the in-plane component of the CDW wavevector, $q_{\text{CDW}}$ at $k_{x,y} = 1/4 a_0$. (**b**) DFT calculations of the phonon dispersion within the harmonic approximation ($T_{\text{el}} = 500$ K). The shaded area highlights the presence of negative (imaginary) phonon frequencies near $q_{\text{CDW}}$. (**c**) Comparison of the harmonic ($T_{\text{el}} = 500$ K) and anharmonic SSCHA ($T_{\text{el}} = 500$ K, $T_{\text{ph}} = 4$ K) calculations close to $q_{\text{CDW}}$. (**d**) Square of the lowest phonon frequency, $\omega^2$ at $q_{\text{CDW}}$ as a function of electronic temperature within the harmonic approximation (squares) and compared to SSCHA (circles). The light pink circle indicates the absolute lowest eigenvalue, $\omega^2_{\text{min}}$ in the entire Brillouin zone for the SSCHA calculation.

been reported in previous theoretical investigations [8, 9]. Instead, it has been shown that by including the effects of anharmonicity, it is possible to achieve estimates that are much closer to reality [8].

11

**Results of the calculation within the anharmonic approximation (SSCHA)**

Following the previous discussion, we next perform calculations including *anharmonic* effects using the stochastic self-consistent harmonic approximation (SSCHA), which has been demonstrated to predict a more realistic transition CDW critical temperature in $1T$-VSe$_2$ at thermal equilibrium [8]. Another advantage of this method is the possibility to define separate temperatures for the electronic ($T_{\text{el}}$) and lattice ($T_{\text{ph}}$) subsystems. This is particularly useful for simulating the situation in ultrafast spectroscopy experiments, where the dominant effect of photoexcitation is to induce a near-instantaneous transiently enhanced electronic temperature, $T_{\text{el}}$ which can be of the order hundreds to thousands of Kelvin, while the thermalization of electrons with phonons occurs on a longer timescale and hence the lattice remains effectively cold within the first few picoseconds.

As discussed previously, we calculated that the critical photocarrier density required to induce the photoinduced CDW-to-normal state transition, as observed in experiments, to be $n_c \approx 1.50 \times 10^{20}$ cm$^{-3}$ which corresponds to an electronic temperature of $T_{\text{el}} = 750$ K. As a conservative estimate for the SSCHA calculations, we used an electronic temperature of $T_{\text{el}} = 500$ K and a lattice temperature of $T_{\text{ph}} = 4$ K to simulate the experimental conditions. The results are shown in Fig. S6c, which compares the calculated phonon dispersion using the *harmonic* and *anharmonic* approximations for $T_{\text{el}} = 500$ K. In the anharmonic case, we find a positive phonon frequency at $q_{\text{CDW}}$ suggesting a melting of the CDW at this temperature, while a weak instability remains ($\omega \leq 10i$ cm$^{-1}$) slightly away from $q_{\text{CDW}}$ towards the $\bar{\text{A}}$-point, highlighted by the shaded area. These results are also reported in Fig. S6d where the circles show the value of $\omega^2$ at $q_{\text{CDW}}$ (dark pink) and the minimum value found across the entire BZ (light pink).

To summarize, the SSCHA results indicate that; (i) we have successfully simulated CDW melting in out-of-equilibrium conditions based on a Fermi-Dirac smearing of the electron occupation similar to the situation induced by the photoexcitation, and (ii) these conditions ($T_{\text{el}} = 500$ K, $T_{\text{ph}} = 4$ K) are very close to the CDW-to-normal state threshold, in full agreement with experiment, since the input to the calculations was based on the experimentally determined critical fluence, $F_c$. Finally, the comparison between the *harmonic* and the *anharmonic* (SSCHA) results highlight the fundamental role of the anharmonic contribution in describing the CDW melting mechanism in photoexcited $1T$-VSe$_2$.



\end{document}